\documentclass{article}

\usepackage{amsfonts,amsmath,amsthm}
\usepackage{subfigure}
\usepackage{multirow}
\usepackage{psfrag}
\usepackage{pifont}
\usepackage{rotating}
\usepackage{threeparttable}
\usepackage{graphicx}

\newtheorem{thm}{Theorem}

\newtheorem{rem}{Remark}

\newcommand{\HHH}{{\mathcal H }}

\newcommand{\PPP}{{\mathcal P }}

\newcommand{\RR}{{\mathbb R}}

\newcommand{\SSS}{{\mathbb  S}}
\newcommand{\CC}{{\mathbb C}}

\newcommand{\dotex}{{\frac{d}{dt}}}

\newcommand{\ket}[1]{\left|#1\right>}
\newcommand{\bra}[1]{\left<#1\right|}

\newcommand{\tr}[1]{\text{Tr}\left[#1\right]}
\newcommand{\her}[2]{\left<#1,#2\right>}

\begin{document}
\bibliographystyle{plain}

\title{Observer-based Hamiltonian identification for quantum systems}
\author{
  Mazyar Mirrahimi\thanks{Corresponding author. INRIA Rocquencourt, Domaine de Voluceau,
 Rocquencourt B.P. 105, 78153 Le Chesnay Cedex, FRANCE.
  Email: mazyar.mirrahimi@inria.fr}
 \and
  Pierre Rouchon\thanks{Ecole des Mines de Paris, Centre Automatique et Syst\`{e}mes,
 60 Bd Saint-Michel, 75272 Paris cedex 06, FRANCE.
  Email: pierre.rouchon@ensmp.fr}
  }

\date{\today}

\maketitle

\begin{abstract}
An observer-based Hamiltonian identification algorithm for quantum
systems is proposed. For the 2-level case an exponential convergence
result based on averaging arguments and some relevant
transformations is provided. The convergence for multi-level cases
is discussed using some heuristic arguments and the relevance of the
method is tested via simulations. Finally, the robustness issue with
respect to non-negligible uncertainties and experimental noises is
also addressed on simulations.
\end{abstract}


\section{Introduction}\label{sec:intro}

The ability of coherent light to manipulate molecular systems at the
quantum scale has been demonstrated both theoretically and
experimentally~\cite{hbref1,scienceref1,hbref4,hbref5,hbref6,rabitz_science2003,
ricebook, scienceref1, rabitz-tracking-95, rabitz-zhu-98,
turinici-maday-03, turinici-control1, lebris-et-al-03,
phan-97,mirrahimi-et-al04, mirrahimi-et-al2-04}. Many of the
procedures, considered in this aim, are based on the possibility to
perform a large number of experiments in a very small time frame.
Thus, the output provided by these experiments can be used to
correct the process and to identify more satisfactory control
fields~\cite{judson-92,phan-97, turinici-control1}.

The ability to rapidly generate a large amount of quantum dynamics
data may  also be used to extract more information about the
possibly unknown parameters of the quantum system itself. For each
test field (i.e., control), there is the possibility of performing
many observations for deducing  information about the system, and
this process can often be carried out at a much faster rate than the
associated numerical simulations of the dynamics. Moreover, the
recent advances in laser technology provide the means for generating
a very large class of test fields for such experiments.

The rapidly developing theory of quantum parameter estimation has
been investigated through different approaches. The
maximum-likelihood methods and the subsequent experiment design
techniques provide a first class of results in this
area~\cite{mabuchi-96,paris-et-al-01,kosut-et-al-03,kosut-et-al-04,
kosut-expdes-04}. The optimal identification techniques via
least-square
criteria's~\cite{geremia-rabitz_02,geremia-rabitz-03,mirrahimi-cocv-05}
and the map inversion techniques~\cite{rabitz-et-al-02} are some
other techniques explored in this area. Finally Kalman filtering
techniques~\cite{geremia-mabuchi-et-al-03,geremia-mabuchi-et-al-04}
have been applied to some atomic magnetometery problems.

However, in general, developing effective identification algorithms
is of a great interest in this domain. The main concerns in quantum
parameter estimation theory are the presence of local minima's for
the optimization problems, sensitivity with respect to the
experimental uncertainties and noises and finally the heavy cost of
computations in formerly developed algorithms.

Before going through the identification and the experiment design
problems, we need to ensure the identifiability of the system. This
issue has been addressed in a recent work~\cite{mirrahimi-cocv-05}
where sufficient assumptions applying the uniqueness of the
inversion result are provided in two relevant settings. A brief
review of an identifiability result needed  for the purpose of this
paper is given in the Appendix~\ref{append:identif}. The
semi-constructive proof in~\cite{mirrahimi-cocv-05} suggests that a
well-chosen control laser field, coupling all the eigenstates of the
free Hamiltonian, would be sufficient to  identify the unknown
parameters.

In~\cite{kosut:ifac04}, a state observer for a known quantum system
is proposed. This observer is then used as a basis for the quantum
parameter estimation applying an iterative search algorithm. The
provided optimization algorithms typically converge toward local
minima's.

Here, in the same direction, we provide an observer-based parameter
estimation algorithm based on techniques derived from adaptive
control theory. In this aim, we will integrate online a generalized
observer including the estimators for both the unknown state and the
unknown parameters of the system.

In the next section, we will present the suggested algorithm on a
simple 2-level system where the unknown parameter to identify is
reduced to a real constant $\theta$ multiplying the dipole moment.
After presenting the system and the estimator, we check the
performance of the method by a first simulation
(Subsection~\ref{ssec:simul}). Subsection~\ref{ssec:design} has for
goal to explain the special choice of the
estimator~\eqref{eq:estpar} and~\eqref{eq:param}. Finally in
Subsection~\ref{ssec:conv}, we provide a detailed proof of a
convergence result.

In Section~\ref{sec:gen}, we extend the observer of
Section~\ref{sec:2lev} to the general case of an $N$-level system.
The convergence of the estimator is discussed using some formal
arguments. The efficiency of the technique is then checked on two
test cases of 3 and 4 dimensions.

Finally, in Section~\ref{sec:robust}, we address the robustness
issue with respect to the measurement and the control noises and
uncertainties. The effect of different kinds of uncertainties and
noises on the identification result has been checked out on
simulations for the 2-level case. Similar simulations for the
multi-level situations give rise to the same kind of robustness.

\section{The 2-level case}\label{sec:2lev}

\subsection{The system and its estimator}\label{ssec:system}

In this section, we consider the 2-level system,
\begin{align}\label{eq:2-lev}
&i\frac{d}{dt}\begin{pmatrix}
    \Psi_1\\
    \Psi_2\\
\end{pmatrix}=(H+u(t)~\theta~\mu)\begin{pmatrix}
    \Psi_1\\
    \Psi_2\\
\end{pmatrix},\qquad \Psi=\begin{pmatrix}
    \Psi_1\\
    \Psi_2\\
\end{pmatrix}\in\CC^2 \notag\\
&y(t)=\her{P\Psi(t)}{\Psi(t)},\\
&H=\frac{\omega}{2}\sigma_z=\begin{pmatrix}
    \omega/2 & 0\\
    0 & -\omega/2\\
\end{pmatrix},
\mu=\sigma_x=\begin{pmatrix}
    0 & 1\\
    1 & 0\\
\end{pmatrix}, P=(1+\sigma_z)/2=\begin{pmatrix}
    1 & 0\\
    0 & 0\\
\end{pmatrix},\notag
\end{align}
where $y$ is the system's output, being the population of the first
eigenstate of $H$. Here $\theta>0$ the dipole moment parameter is
supposed to be unknown. The goal of this section is to identify this
unknown parameter.

In the density matrix language, this same system can be written as,
\begin{equation}\label{eq:density}
\frac{d}{dt}\rho(t)=-i~[H+u(t)~\theta~\mu,\rho(t)],\qquad
y(t)=\tr{P\rho(t)},
\end{equation}
where $\rho(t)=\Psi\Psi^\dag \in \CC^{2\times 2}$ is the projection
matrix on the wavefunction $\Psi(t)$.

We consider the laser field $u(t)$ to be in the resonant regime with
respect to the natural frequency of the system (being the difference
between the eigenvalues of the Hamiltonian $H$, $-\omega/2$ and
$\omega/2$):
$$
u(t)=A(t)\cos(\omega t),
$$
where $A(t)$ is a slowly variable modulation of the amplitude. Here,
for simplicity sakes, we consider a constant amplitude $A(t)\equiv
A>0$.

In this paper, we propose the following observer-based parameter
estimator,
\begin{alignat}{2}
\frac{d}{dt}\hat\rho&= -i[H+\hat\theta~u(t)\mu,\hat\rho] + \Gamma
(y(t)-\hat y(t)) \left(P\hat\rho+\hat\rho
P-2\tr{P\hat\rho}\hat\rho\right),\label{eq:estpar}\\
\hat y &=\tr{P\hat \rho},\notag\\
\frac{d}{dt}\hat\theta &=-i\gamma~u(t)\tr{P[\mu,\hat\rho]}(y(t)-\hat
y(t)),\label{eq:param}
\end{alignat}
where $\gamma$ and $\Gamma$ are positive real design parameters. As
we will see in Subsection~\ref{ssec:conv}, in theory, we need the
following validity domain in order to ensure the convergence toward
the true parameter $\theta$:
$$
 \Gamma\ll A\theta
 ,\quad A\theta\ll \omega
  ,\quad \gamma \ll \theta.
$$
However, the simulations of the next subsection show that these
restrictions can be much relaxed in practice. In particular, one can
use instead of~\eqref{eq:estpar} and~\eqref{eq:param}, the average
estimator of Remark~\ref{rem:match}, where the high Bohr frequency
is removed and that does not require the precise knowledge of the
transition frequency.

\subsection{Simulations}\label{ssec:simul}

Before explaining the choice of this identification algorithm and
going through the technicalities of a convergence proof, let us
check the efficiency of the algorithm in a simulation. \small
\begin{figure}[h]
\begin{center}
\includegraphics[width=9 cm]{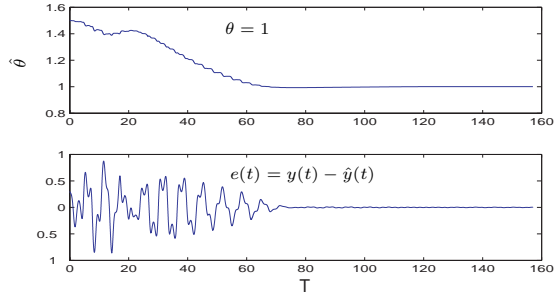}
\caption{The first figure shows the convergence of the parameter
estimator $\hat\theta$ towards its true value $1$; the second shows
the convergence of the error term $e(t)$ toward
0.}\label{fig:2lev-1}
\smallskip
\end{center}
\end{figure}
\normalsize Here, we assume the unknown parameter $\theta$, the
frequency $\omega$, the constant $\Gamma$ and the laser amplitude
$A$ to be all equal to 1. The constant $\gamma$ in~\eqref{eq:param}
is chosen to be $\gamma=0.1$. The real system $\rho$ and the
estimator $\hat \rho$ are respectively initialized at
$\rho(0)=\Psi_0\Psi_0^\dag$ and
$\hat\rho(0)=\hat\Psi_0\hat\Psi_0^\dag$ where
$\Psi_0=(1/\sqrt{2},1/\sqrt{2})^T$ and
$\hat\Psi_0=(1/\sqrt{5},2/\sqrt{5})^T$. Finally the parameter
estimator $\hat\theta$ is initialized at $\hat\theta(0)=1.5$.

The simulations of Figure~\ref{fig:2lev-1} show the result for the
algorithm presented above. The simulation time $T=50\pi$ represents
25 times the natural period, at which the system without control
oscillates. As it can be seen the above algorithm ensures the
convergence of the parameter estimator $\hat\theta(t)$ toward the
unknown parameter $\theta=1$.

\subsection{Estimator design}\label{ssec:design}

In this subsection, we will explain the particular choice of the
observer-based estimator~\eqref{eq:estpar} and~\eqref{eq:param}.

Starting from the physical system~\eqref{eq:2-lev}, whenever the
parameter $\theta$ is known, an intuitive observer (similar to the
one proposed in~\cite{kosut:ifac04}) can be given as follows:
\begin{equation}\label{eq:obs}
\frac{d}{dt}\tilde\Psi=-i(H+\theta~u(t)\mu)\tilde\Psi+\Gamma
(y(t)-\tilde y(t)) P\tilde\Psi,
\end{equation}
where $\tilde y(t)=\her{P \tilde\Psi (t)}{\tilde\Psi (t)}$ and
$\Gamma$ is a positive constant.

The non-conservative wave-functional equation~\eqref{eq:obs}
($\|\tilde\Psi(t)\|$ does not remain constant equal to 1)  can be
written in the density matrix formulation as follows:
\begin{equation}\label{eq:estpar1}
\frac{d}{dt}\tilde\rho= -i [H+\theta~u(t)\mu,\tilde\rho] + \Gamma
(y(t)-\tilde y(t)) \left(P\tilde\rho+\tilde\rho P\right),
\end{equation}
where $\tilde\rho=\tilde\Psi\tilde\Psi^\dag$ and $\tilde
y=\tr{P\tilde\rho}$.

Note that unlike the conservative Schr{\"o}dinger
equation~\eqref{eq:density}, the observer
equation~\eqref{eq:estpar1} does not conserve the trace of the
density matrix $\tilde\rho$. However, as we will see in
Subsection~\ref{ssec:conv}, enforcing the observer to keep the same
geometrical structure as the main system simplifies considerably the
convergence proof. In this aim we define a normalized density matrix
given by $\hat\rho=\tilde\rho/\tr{\tilde\rho}$. This normalized
observer state verifies a conservative equation of the following
form
\begin{equation}\label{eq:estpar2}
\frac{d}{dt}\hat\rho= -i[H+\theta~u(t)\mu,\hat\rho] + \Gamma
(y(t)-\hat y(t)) \left(P\hat\rho+\hat\rho
P-2\tr{P\hat\rho}\hat\rho\right),
\end{equation}
where $\hat y=\tr{P\hat\rho}$.

Whenever, the parameter $\theta$ is unknown, one might consider it
as a part of the system's state to be estimated. In this aim, we
replace the parameter $\theta$ in~\eqref{eq:estpar2} by $\hat\theta$
to be estimated (so that we obtain~\eqref{eq:estpar}). The question
becomes now to provide an evolution equation for $\hat\theta$
ensuring the convergence toward the true parameter $\theta$.

Consider the Lyapunov function
\begin{equation}\label{eq:lyap2}
V(t,\hat\rho,\hat\theta)=\frac{1}{2}(y(t)-\hat
y(t))^2+\frac{1}{2\gamma}(\theta-\hat\theta)^2,
\end{equation}
where $\gamma$ is a small enough positive constant. Deriving with
respect to time, we have:
\begin{multline*}
\frac{d}{dt}V=\left(i\hat\theta(t)~u(t)\tr{P[\mu,\hat\rho(t)]}-
i\theta~u(t)\tr{P[\mu,\rho(t)]}\right)e(t)\\
+\frac{1}{\gamma}\dot{\hat\theta}(\hat\theta(t)-\theta)-2\Gamma
|e(t)|^2\tr{P\hat \rho}(1-\tr{P\hat \rho}),
\end{multline*}
where $e(t)=y(t)-\hat y(t)$. Let us take
\begin{equation*}
\frac{d}{dt}\hat\theta=-i\gamma~u(t)\tr{P[\mu,\hat\rho]}e(t).
\end{equation*}
Then we have:
\begin{equation}\label{eq:der}
\frac{d}{dt}V = (i\theta~u(t)~\tr{P[\mu,\hat\rho-\rho]})-2\Gamma
|e(t)|^2\tr{P\hat \rho}(1-\tr{P\hat \rho}),
\end{equation}
While the last term in~\eqref{eq:der} is always negative, the
situation has no reason to be the same for the first one. We will
see however that under certain circumstances this first term can be
neglected and the Lyapunov function represents a decreasing behavior
in average.

\begin{figure}[h]
\begin{center}
\includegraphics[width=10 cm]{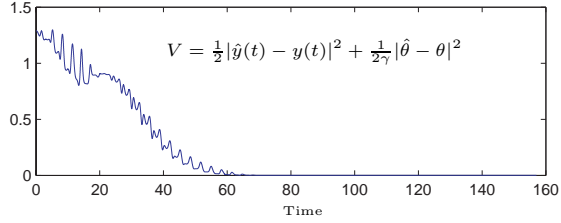}
\caption{The positive definite function $V$ defined
in~\eqref{eq:lyap2} with $\gamma=.1$; the function decreases in
average and ends up converging toward 0. } \label{fig:2lev-2}
\end{center}
\end{figure}

The Figure~\ref{fig:2lev-2} illustrates the evolution of this
Lyapunov function for the simulations of the last subsection.

\subsection{Convergence analysis}\label{ssec:conv}

One has the following convergence result:
\begin{thm}\label{thm:conv}
Consider the 2-levels system describes by
\begin{equation}\label{dyn:eq}
 \left\{
 \begin{aligned}
 \frac{d}{dt}\rho(t)& =-i~[H+u(t)\theta\mu,\rho(t)]
 \\
y(t)& =\tr{P\rho(t)}
 \end{aligned}
 \right.
\end{equation}
where $\rho(t)$ is the density matrix, $H=\frac{\omega}{2}
\sigma_z$, $\mu=\sigma_x$, $P=(1+\sigma_z)/2$. The goal is to
estimate $\theta
>0 $ and $\rho$ via the measure $y$ and the knowledge of $\omega$. Assume that the estimations
$\hat \theta$ and $\hat \rho$ obey the following dynamics (a
nonlinear filter of $y$):
\begin{equation}\label{dyne:eq}
 \left\{
 \begin{aligned}
\frac{d}{dt}\hat\rho&= -i[H+\hat\theta~u(t)\mu,\hat\rho] + \Gamma
(y(t)-\tr{P\hat \rho}) \left(P\hat\rho+\hat\rho
P-2\tr{P\hat\rho}\hat\rho\right)
\\
\frac{d}{dt}\hat\theta
&=-i\gamma~u(t)\tr{P[\mu,\hat\rho]}(y(t)-\tr{P\hat \rho}).
 \end{aligned}
 \right.
\end{equation}
where  $\Gamma$ and $\gamma$ are two positive  gains. Assume that
$u(t)=A\cos(\omega t)$ where $A$ is a  constant amplitude. Assume
that the  $\rho(0)$ does not belong to $\left\{\frac{1-\sigma_x}{2}
,\frac{1+\sigma_x}{2} \right\}$.  Then exist $\epsilon
>0$ and $\eta
>0$, such that, for any design parameters $(A,\Gamma,\gamma)$
satisfying
\begin{equation}\label{gain:eq}
 0<\Gamma\leq \epsilon A\theta
 ,\quad 0< A\theta\leq \epsilon\omega
  ,\quad 0<  \gamma \leq \epsilon \theta
\end{equation}
and  any initial conditions $(\hat\rho_0,\hat\theta_0)$ satisfying
($\hat\rho_0$ corresponds to a pure state, $\hat\rho_0$ is symmetric
positive matrix of trace one and $\hat\rho^2_0=\hat\rho_0$):
$$
\tr{(\hat\rho_0-\rho(0))^2}\leq \eta, \quad |\hat\theta_0
-\theta|\leq \eta,
$$
the estimates $(\hat\rho,\hat\theta)$ converge:
$$
\lim_{t\mapsto+\infty} (\hat\rho(t)-\rho(t))=0, \quad
\lim_{t\mapsto+\infty} \hat\theta(t)=\theta .
$$
Moreover, this convergence is exponential (locally): it is robust to
small modeling  and measurements errors.
\end{thm}

\begin{rem}\label{rem:match}
The fact that $\rho(0)$ must be different from $(1\pm\sigma_x)/2$ is
 not a severe  limitation in practice since an alternative
 to~\eqref{dyne:eq}  will be the first  averaged
 system~\eqref{dynA1:eq}. It provides a more realistic estimator
 since it does not depends on  $\omega$ the Bohr transition frequency that
 is large in general:
$$
 \left\{
 \begin{aligned}
   \dotex\hat\xi =&
      -i \frac{A\hat\theta}{2}
       [\sigma_x,\hat\xi]
        +\frac{\Gamma}{4} \left(y(t)-\tr{\sigma_z\hat\xi}\right)
        \left(\sigma_z\hat\xi+\hat\xi \sigma_z-2\tr{\sigma_z\hat\xi}\hat\xi\right)
\\
    \dotex \hat\theta =&
 \frac{\gamma A}{4}
       \tr{\sigma_y\hat\xi} \left(y(t)-\tr{\sigma_z\hat\xi}\right)
       .
 \end{aligned}
 \right.
$$
More-over if the laser frequency $\omega_r$  does not perfectly
match the resonance condition $u(t)=A\cos(\omega_r(t))$,  the
averaged estimator  reads \small
$$
 \left\{
 \begin{aligned}
   \dotex\hat\xi =&
   -i\frac{\omega-\omega_r}{2}[\sigma_z,\hat\xi]
      -i \frac{A\hat\theta}{2}
       [\sigma_x,\hat\xi]
        +\frac{\Gamma}{4} \left(y(t)-\tr{\sigma_z\hat\xi}\right)
        \left(\sigma_z\hat\xi+\hat\xi \sigma_z-2\tr{\sigma_z\hat\xi}\hat\xi\right)
\\
    \dotex \hat\theta =&
 \frac{\gamma A}{4}
       \tr{\sigma_y\hat\xi} \left(y(t)-\tr{\sigma_z\hat\xi}\right)
       .
 \end{aligned}
 \right.
$$\normalsize
When $|\omega-\omega_r| \ll A\theta$, the second averaged system is
then identical to~\eqref{dynA2:eq} and thus convergence is ensured.
Thus a precise knowledge of $\omega$ is not necessary for estimating
$\theta$.
\end{rem}

\begin{rem}\label{rem:unnorm}
The simulations show that the un-normalized observer
equation~\eqref{eq:obs} (and the parameter estimator found by
adapting this equation to the Lyapunov function~\eqref{eq:lyap2})
would be sufficient to ensure the convergence. However, as we will
see in the proof below, the trace conservation for the normalized
equation simplifies considerably the analysis  since $\hat\rho$
remains on the Bloch sphere.
\end{rem}

The  proof is based on two successive averaging: the first one
relies on the resonant control $u(t)$, removes the laser frequency
$\omega$  and yields to~\eqref{dynA1:eq}; the second one eliminates
the Rabi frequency $A\theta$ ($A\theta \ll \omega$) and
provides~\eqref{dynA2:eq}. This second averaged system is proved to
be locally exponentially convergent. This implies the exponential
convergence of the first averaged system~\eqref{dynA1:eq} and of the
original system~\eqref{dyne:eq}.

\begin{proof}
Notice first that~\eqref{dyne:eq} reads
$$
  \left\{
 \begin{aligned}
  \frac{d}{dt}\hat\rho&= -i\left[\frac{\omega}{2}\sigma_z+\hat\theta~u(t)\sigma_x,\hat\rho\right] +
  \frac{\Gamma}{4} \tr{\sigma_z(\rho-\hat \rho)}
   \left(\sigma_z\hat\rho+\hat\rho
  \sigma_z-2\tr{\sigma_z\hat\rho}\hat\rho\right)
  \\
  \frac{d}{dt}\hat\theta
  &=\frac{\gamma u(t)}{2}\tr{\sigma_y\hat\rho}
  \tr{\sigma_z(\rho-\hat \rho)}.
 \end{aligned}
 \right.
$$
Since $\tr{\hat\rho(0)}=1$ and $\tr{\hat\rho^2(0)}=1$, we have
$\tr{\hat\rho(t)} \equiv 1$ and $\tr{\hat\rho^2(t)}\equiv 1$. Thus
$\hat\rho$ remains a positive
 matrix associated to a pure state $\ket{\hat\psi}\in \CC^2$ of length one:
 $\hat\rho = \ket{\hat\psi} \bra{\hat\psi}$:
$$
  \hat\rho = \frac{1
     + \tr{\sigma_x\hat\rho} \sigma_x
     + \tr{\sigma_y\hat\rho} \sigma_y
     + \tr{\sigma_z\hat\rho} \sigma_z
 }{2}
$$
where the  Bloch vector of components
$(\tr{\sigma_x\hat\rho},\tr{\sigma_y\hat\rho},\tr{\sigma_z\hat\rho})$
remains on the unit sphere $\SSS^2$.

 The resonance assumption for the control field allows us to reduce
the system by averaging and removing highly oscillating terms of
frequency $\omega$. Consider the following time-dependent change of
variables
$$
 \xi =e^{i \frac{\omega t\sigma_z}{2}}
 \rho e^{-i \frac{\omega t\sigma_z}{2}}
   ,\quad
 \hat\xi=e^{i \frac{\omega t\sigma_z}{2}}
 \hat\rho e^{-i \frac{\omega t\sigma_z}{2}}.
$$
Since $[\sigma_z,P]=0$, we have:
\begin{align*}
   \dotex \xi =& -i A\theta
   \cos(\omega t)
   \left[e^{i \frac{\omega t\sigma_z}{2}}\sigma_x
   e^{-i \frac{\omega t\sigma_z}{2}},\xi\right]
\\
   \dotex\hat\xi =&
      -i A\hat\theta \cos (\omega t)
       \left[e^{i \frac{\omega t\sigma_z}{2}}\sigma_x
       e^{-i\frac{\omega t\sigma_z}{2}},\hat\xi\right]
       \\
       &\quad
        +\frac{\Gamma}{4}  \tr{\sigma_z(\xi-\hat\xi)}
        \left(\sigma_z\hat\xi+\hat\xi \sigma_z-2\tr{\sigma_z\hat\xi}\hat\xi\right)
\\
    \dotex \hat\theta =&
     \frac{\gamma A}{2} \cos(\omega t)
       \tr{e^{i \frac{\omega t\sigma_z}{2}}\sigma_y
       e^{-i \frac{\omega t\sigma_z}{2}}\hat\xi}
      \tr{\sigma_z(\xi-\hat\xi)}
       .
\end{align*}
But
$$
    e^{i\frac{\omega t}{2}\sigma_z}
      \sigma_x e^{-i\frac{\omega t}{2} \sigma_z} =
    e^{i\omega t\sigma_z}
      \sigma_x =
   \cos(\omega t) \sigma_x - \sin(\omega t ) \sigma_y.
$$
and
$$
    e^{i\frac{\omega t}{2}\sigma_z}
      \sigma_y e^{-i\frac{\omega t}{2} \sigma_z} =
    e^{i\omega t\sigma_z}
      \sigma_y =
   \cos(\omega t) \sigma_y + \sin(\omega t ) \sigma_x.
$$
Take then the averaged system where the rapidly oscillating terms
(associated to  $\sin(2\omega t)$ or  $\cos(2\omega t)$) are
removed:
\begin{equation}\label{dynA1:eq}
 \left\{
 \begin{aligned}
   \dotex \xi =&
       -i \frac{A\theta}{2} [\sigma_x,\xi]
\\
   \dotex\hat\xi =&
      -i \frac{A\hat\theta}{2}
       [\sigma_x,\hat\xi]
        +\frac{\Gamma}{4} \tr{\sigma_z(\xi-\hat\xi)}
        \left(\sigma_z\hat\xi+\hat\xi \sigma_z-2\tr{\sigma_z\hat\xi}\hat\xi\right)
\\
    \dotex \hat\theta =&
 \frac{\gamma A}{4}
       \tr{\sigma_y\hat\xi} \tr{\sigma_z(\xi-\hat\xi)}
       .
 \end{aligned}
 \right.
\end{equation}
Consider now the new  variables,
$$
  \zeta=e^{i \frac{ A\theta t\sigma_x}{2}} \xi
    e^{-i \frac{ A\theta t\sigma_x}{2}}
    ,\quad
  \hat\zeta=e^{i \frac{ A\theta t\sigma_x}{2}} \hat\xi
    e^{-i \frac{ A\theta t\sigma_x}{2}}
    .
$$
Then~\eqref{dynA1:eq} reads:
\begin{equation}\label{dynA11:eq}
 \begin{aligned}
   \dotex \zeta &= 0
\\
   \dotex\hat\zeta &=
      -i \frac{A(\hat\theta-\theta)}{2}
       [\sigma_x,\hat\zeta]
       \\
       &\quad
        +\frac{\Gamma}{4} \tr{e^{i \frac{A\theta t\sigma_x}{2}}
   \sigma_z e^{-i \frac{A\theta t\sigma_x}{2}}(\zeta-\hat\zeta)}
   \\&\qquad\qquad
         \left(
          e^{i \frac{A\theta t\sigma_x}{2}}
   \sigma_z e^{-i \frac{A\theta t\sigma_x}{2}}\hat\zeta
          +\hat\zeta e^{i \frac{A\theta t\sigma_x}{2}}
   \sigma_z e^{-i \frac{A\theta t\sigma_x}{2}}
          -2\tr{e^{i \frac{A\theta t\sigma_x}{2}}
   \sigma_z e^{-i \frac{A\theta t\sigma_x}{2}}\hat\zeta}\hat\zeta\right)
\\
    \dotex \hat\theta &=
     \frac{\gamma A}{4}
       \tr{e^{i \frac{A\theta t\sigma_x}{2}}
   \sigma_y e^{-i \frac{A\theta t\sigma_x}{2}}\hat\zeta}
      \tr{e^{i \frac{A\theta t\sigma_x}{2}}
   \sigma_z e^{-i \frac{A\theta t\sigma_x}{2}}(\zeta-\hat\zeta)}
       .
\end{aligned}
\end{equation}
But we have
\begin{align*}
 e^{i \frac{A\theta t\sigma_x}{2}}
   \sigma_y e^{-i \frac{A\theta t\sigma_x}{2}}= &
 \cos(A\theta t)\sigma_y - \sin(A\theta t)\sigma_z
 \\
 e^{i \frac{A\theta t\sigma_x}{2}}
   \sigma_z e^{-i \frac{A\theta t\sigma_x}{2}} =&
  \sin(A\theta t)\sigma_y+\cos(A\theta t)\sigma_z
\end{align*}
Let us consider now the secular terms in~\eqref{dynA11:eq}, i.e.,
terms without  the rapidly oscillating factor $\sin(2A\theta t)$ or
$\cos(2A\theta t)$. The secular term  in
\begin{multline*}
\tr{e^{i \frac{A\theta t\sigma_x}{2}}
   \sigma_z e^{-i \frac{A\theta t\sigma_x}{2}}(\zeta-\hat\zeta)}
   \\
         \left(
          e^{i \frac{A\theta t\sigma_x}{2}}
   \sigma_z e^{-i \frac{A\theta t\sigma_x}{2}}\hat\zeta
          +\hat\zeta e^{i \frac{A\theta t\sigma_x}{2}}
   \sigma_z e^{-i \frac{A\theta t\sigma_x}{2}}
          -2\tr{e^{i \frac{A\theta t\sigma_x}{2}}
   \sigma_z e^{-i \frac{A\theta t\sigma_x}{2}}\hat\zeta}\hat\zeta\right)
\end{multline*} is sum of two quantities
\begin{multline*}
      \frac{1}{2}\tr{\sigma_y(\zeta-\hat\zeta)}
         \left(
          \sigma_y\hat\zeta
          +\hat\zeta \sigma_y
          -2\tr{\sigma_y\hat\zeta}\hat\zeta\right)
          \\
          +
      \frac{1}{2} \tr{\sigma_z(\zeta-\hat\zeta)}
         \left(
          \sigma_z\hat\zeta
          +\hat\zeta \sigma_z
          -2\tr{\sigma_z\hat\zeta}\hat\zeta\right)
\end{multline*}
The secular term for
 $\tr{e^{i \frac{A\theta t\sigma_x}{2}}
   \sigma_y e^{-i \frac{A\theta t\sigma_x}{2}}\hat\zeta}
   \tr{e^{i \frac{A\theta t\sigma_x}{2}}
   \sigma_z e^{-i \frac{A\theta t\sigma_x}{2}}(\zeta-\hat\zeta)}$ is
$$
\frac{1}{2}
\left(\tr{\sigma_y\hat\zeta}\tr{\sigma_z(\zeta-\hat\zeta)}
      -
      \tr{\sigma_z\hat\zeta}\tr{\sigma_y(\zeta-\hat\zeta)}
      \right)
$$
Thus the averaged system associated to~\eqref{dynA1:eq} reads:
\begin{equation}\label{dynA2:eq}
\left\{
 \begin{aligned}
   \dotex \zeta &= 0
\\
   \dotex\hat\zeta &=
      -i \frac{ A(\hat\theta-\theta)}{2}
       [\sigma_x,\hat\zeta]
  \\
  &\qquad
  +
      \frac{\Gamma}{8}\tr{\sigma_y(\zeta-\hat\zeta)}
         \left(
          \sigma_y\hat\zeta
          +\hat\zeta \sigma_y
          -2\tr{\sigma_y\hat\zeta}\hat\zeta\right)
  \\
  &\qquad
  +
      \frac{\Gamma}{8} \tr{\sigma_z(\zeta-\hat\zeta)}
         \left(
          \sigma_z\hat\zeta
          +\hat\zeta \sigma_z
          -2\tr{\sigma_z\hat\zeta}\hat\zeta\right)
\\
    \dotex \hat\theta &=
    \frac{\gamma A}{8}
\left(\tr{\sigma_y\hat\zeta}\tr{\sigma_z(\zeta-\hat\zeta)}
      -
      \tr{\sigma_z\hat\zeta}\tr{\sigma_y(\zeta-\hat\zeta)}
      \right)       .
\end{aligned}
\right.
\end{equation}
 Consider now the
following Lyapounov function:
$$
  V(\hat\zeta,\hat\theta)
  =
  \frac{1}{2} \tr{\sigma_y (\hat \zeta-\zeta)}^2 +
  \frac{1}{2} \tr{\sigma_z (\hat\zeta-\zeta)}^2
  +\frac{4}{\gamma } (\hat\theta -\theta)^2
  .
$$
One has
\begin{multline*}
  \frac{4}{\Gamma}\dotex V =
     \left(
       \tr{\sigma_y(\hat\zeta-\zeta)}\tr{\sigma_y\hat\zeta}
       +
       \tr{\sigma_z(\hat\zeta-\zeta)}\tr{\sigma_z\hat\zeta}
      \right)^2
      \\
     - \tr{\sigma_y(\hat\zeta-\zeta)}^2
     - \tr{\sigma_z(\hat\zeta-\zeta)}^2.
\end{multline*}
We have (Cauchy-Schwartz inequality)
$$
\frac{4}{\Gamma}\dotex V \leq \left(\tr{\sigma_y\hat\zeta}^2 +
\tr{\sigma_z\hat\zeta}^2
-1\right)\left(\tr{\sigma_y(\hat\zeta-\zeta)}^2
     + \tr{\sigma_z(\hat\zeta-\zeta)}^2\right)
$$
Since $\tr{\sigma_y\hat\zeta}^2 + \tr{\sigma_z\hat\zeta}^2\leq 1$,
$\dotex V \leq 0$. When $\dotex V=0$ we have
\begin{itemize}
\item either
$$ \tr{\sigma_y(\hat\zeta-\zeta)}^2
     + \tr{\sigma_z(\hat\zeta-\zeta)}^2 = 0
     .
$$ This means that $\tr{\sigma_x\hat\zeta}=\pm\tr{\sigma_x\zeta}$:
\begin{equation*}
\hat\zeta=\zeta_{\pm}=\frac{I\pm\tr{\sigma_x\zeta}\sigma_x+\tr{\sigma_y\zeta}\sigma_y+\tr{\sigma_z\zeta}\sigma_z}{2}.
\end{equation*}
We can use here Lasalle invariance  principle since $\hat\zeta$
evolves on a compact manifold and $V$ is infinite when $\hat\theta$
is infinite. Since  $\hat\zeta$ is constant, we have $\dotex
\hat\zeta= 0$ and thus $(\hat\theta-\theta)[\sigma_x,\zeta]=0$.
Since $\rho(0)\neq (1\pm\sigma_x)/2$,
$\zeta(t)=\zeta(0)\approx\rho(0) \neq (1\pm\sigma_x)/2$ and  thus
$[\sigma_x,\zeta]\neq 0$. This implies that $\hat\theta=\theta$. A
simple inspection shows that $(\zeta_{-},\theta)$ is an unstable
equilibria for $(\hat\zeta,\hat\theta)$ and
$(\zeta_{+}=\zeta,\theta)$ is an asymptotically and exponentially
stable one (use on the first order approximation the same Lyapounov
function and the invariance principle that proves the asymptotic
stability of the first variation, and thus its exponentially
stability).

\item or (equality case for the Cauchy-Schwartz inequality)
$$
 \tr{\sigma_y\hat\zeta}=\alpha\tr{\sigma_y\zeta}
 , \quad
 \tr{\sigma_z\hat\zeta}=\alpha\tr{\sigma_z\zeta},
$$
for some real $\alpha$ and
$$
\tr{\sigma_y\hat \zeta}^2+\tr{\sigma_z\hat \zeta}^2=1.
$$
Thus we have
\begin{equation*}
\hat
\zeta=\zeta_{\pm}=\frac{I\pm\left(\frac{\tr{\sigma_y\zeta}}{\sqrt{\tr{\sigma_y\zeta}^2+\tr{\sigma_z\zeta}^2}}\sigma_y+\frac{\tr{\sigma_z\zeta}}{\sqrt{\tr{\sigma_y\zeta}^2+\tr{\sigma_z\zeta}^2}}\sigma_z\right)}{2}.
\end{equation*}
Moreover using LaSalle invariance principle and developing
$d\hat\zeta/dt=0$, we obtain $\hat\theta=\theta$. One can easily see
that $(\hat\zeta=\zeta_\pm,\hat \theta=\theta)$ are also equilibrium
points of the system.

However, note that we are looking for a local result. For initial
state $(\hat\zeta_0,\hat\theta_0)$ near enough to the initial state
$(\zeta,\theta)$, we have
$$
V(\hat\zeta_0,\hat\theta_0)<\epsilon\ll 1.
$$
As the Lyapunov function $V$ keeps decreasing and by choosing
$\epsilon$ to be small enough
($\epsilon<\frac{1}{2}\left(1+\sqrt{\tr{\sigma_y\zeta}^2+\tr{\sigma_z\zeta}^2}\right)^2$),
the state $(\hat \zeta,\hat\theta)$ can not reach the equilibrium
states $(\zeta_{-},\theta)$ where we have:
\begin{equation*}
V(\zeta_-,\theta)=\frac{1}{2}\left(1+\sqrt{\tr{\sigma_y\zeta}^2+\tr{\sigma_z\zeta}^2}\right)^2>\epsilon.
\end{equation*}
Concerning the equilibria $(\zeta_+,\theta)$, we still have two
situations:
\begin{enumerate}
  \item either $\tr{\sigma_x\zeta}=0$, in which case $\zeta_+=\zeta$
  and we have the convergence as we wanted to prove in the theorem.
  \item or $\tr{\sigma_x\zeta}\neq 0$ which implies that $\sqrt{\tr{\sigma_y\zeta}^2+\tr{\sigma_z\zeta}^2}<1$.
  In this case we choose
  $$
    V(\hat\zeta_0,\hat\theta_0)<\epsilon\ll 1,
  $$
  with
  $0<\epsilon<\frac{1}{2}\left(1-\sqrt{\tr{\sigma_y\zeta}^2+\tr{\sigma_z\zeta}^2}\right)^2$.
  As the Lyapunov function $V$ keeps decreasing, the state $(\hat \zeta,\hat\theta)$ can not reach the equilibrium
  states $(\zeta_{+},\theta)$ where we have:
  \begin{equation*}
    V(\zeta_+,\theta)=\frac{1}{2}\left(1-\sqrt{\tr{\sigma_y\zeta}^2+\tr{\sigma_z\zeta}^2}\right)^2>\epsilon.
  \end{equation*}
\end{enumerate}

\end{itemize}
This asymptotic analysis based on the above Lyapounov function and
Lasalle invariance principle shows that the steady-state
$(\hat\zeta,\hat\theta)=(\zeta,\theta)$ of the average
system~\eqref{dynA2:eq} is locally exponentially stable.

The existence of $\epsilon$ and $\eta$ results from  a  classical
lemma concerning the averaging techniques (cf.~\cite{khalil-book},
Page 333, Theorem 8.3): if the average system admits an
exponentially stable steady-state that is also an equilibrium of the
original system, then this steady-state  is also exponentially
stable for the original system.

Consider first~\eqref{dynA2:eq}:
\begin{itemize}
  \item It admits $(\zeta,\theta)$ as exponentially stable
  steady-state.
  \item it corresponds to the averaging of~\eqref{dynA11:eq} when
$A\theta \gg \Gamma, \gamma A$.
\end{itemize}
Thus the state $(\hat\zeta,\hat\theta)$ of~\eqref{dynA11:eq}
converge exponentially towards $(\zeta,\theta)$. Since
$$
   \hat\xi =
   e^{-i \frac{ A\theta t\sigma_x}{2}} \hat\xi
    e^{+i \frac{ A\theta t\sigma_x}{2}}
   ,\quad
   \xi =
   e^{-i \frac{ A\theta t\sigma_x}{2}} \xi
    e^{+i \frac{ A\theta t\sigma_x}{2}}
$$
 $\hat \xi-\xi$ and $ \hat\theta-\theta$  converge
exponentially towards $0$. when $\omega \gg A\theta$, a similar
argument between~\eqref{dyne:eq} and~\eqref{dynA1:eq} yields to the
local exponential convergence of $\hat\rho -\rho$ and $\hat\theta
-\theta$ towards $0$.

\end{proof}

\begin{rem}\label{rem:local}
Theorem~\ref{thm:conv} provides a local convergence result for the
estimator equations given by~\eqref{dyne:eq}. The simulations,
however, show a much stronger global convergence behavior. Notice
that we cannot prove directly that~\eqref{dynA1:eq} is stable,
although  the following function
$$
 V(\hat\xi - \xi,\hat\theta-\theta)
 =
 \frac{1}{2} \tr{\sigma_x(\hat\xi -\xi)}^2 +
 \frac{2}{\gamma} (\hat\theta -\theta)^2
$$
is decreasing in average since
$$
 \dotex V
  = -\frac{\Gamma}{2} \tr{\sigma_z(\hat\xi - \xi)}^2
  \left(1-\tr{\sigma_z\hat\xi}^2\right) +
  A\theta\tr{\sigma_y(\hat\xi-\xi)}
  .
$$
The first term of the left hand side is always negative whereas the
second one admits a zero average. Thus in average, $V$ is a
decreasing function of time.

\end{rem}

\section{The general case}\label{sec:gen}

In this section, we extend the above identification algorithm to the
general case of a multi-level system. Before going through this
extension, we need to address the non-trivial identifiability
problem for the multi-level case. The Appendix~\ref{append:identif}
(based on the result of~\cite{mirrahimi-cocv-05}) provides
sufficient assumptions ensuring the identifiability.

\subsection{Formal extension}\label{ssec:sysgen}
Consider the $N$-levels system, $(\ket{j})_{1\leq j\leq N}$,
described by the density matrix $\rho$ that obey the following
dynamics (we assume here the assumptions \textbf{A1},\textbf{A2} and
\textbf{A3} of the Appendix~\ref{append:identif} to be satisfied):
\begin{equation}\label{dynN:eq}
 \left\{
 \begin{aligned}
  \frac{d}{dt}\rho & = -i [H+u(t)\mu,\rho]
  \end{aligned}
 \right.
\end{equation}
where
\begin{itemize}

  \item $H=\sum_{j=1}^N \omega_j\ket{j}\bra{j}$ is the free
  Hamiltonian  with $\omega_j$ real and satisfying $|\omega_l-\omega_k|\neq
|\omega_{l^\prime}-\omega_{k^\prime}|$ for any distinct couples
$(l,k)$ and $(l^\prime,k^\prime)$.

   \item $\mu=\sum_{1\leq l < k\leq N}
   \theta_{lk}\left(\ket{k}\bra{l}+\ket{l}\bra{k}\right)$ where
   $\theta_{lk}$  are the parameters to identify
   \item the electromagnetic  field is represented by the scalar input $u(t)\in\RR$
\end{itemize}
We assume that
$$
y_j(t)=\tr{P_j\rho(t)},\quad P_j=\ket{j}\bra{j},\quad j=1,2,...,N
$$
are the  measured outputs. The goal is to estimate the coefficient
$\theta_{lk}$. The $\omega_j$ are known.

The estimator~\eqref{dyne:eq} admits then the following
generalization $(1\leq l < k \leq N)$:
\begin{equation}\label{dynNe:eq}
 \left\{
 \begin{aligned}
   \dotex\hat\rho &=-i[H+u(t)\hat\mu,\hat\rho]
    +\Gamma\sum_{j=1}^N
      \left(y_j(t)-\tr{P_j\hat\rho}\right)
      \left(P_j\hat\rho+\hat\rho P_j-2\tr{P_j\hat\rho}\hat\rho\right)
      \\
   \dotex \hat\theta_{lk} &=
-i\gamma_{lk}~u(t)\sum_{j=1}^N
  \tr{P_j\left[\ket{l}\bra{k}+\ket{k}\bra{l},\hat\rho\right]}
  \left(y_j(t)-\tr{P_j\hat\rho}\right)
 \end{aligned}
 \right.
\end{equation}
where $\hat \mu=\sum_{1\leq l<k\leq N}
   \hat
   \theta_{lk}\left(\ket{l}\bra{k}+\ket{k}\bra{l}\right)
   $,  $\Gamma>0$ and $\gamma_{lk}>0$ are design parameters.
   Notice that $\hat\rho$ remains a projector if its initial condition
is also a projector.

Set $u(t)= \sum_{1\leq l<k\leq N} A_{lk}\cos(\omega_{lk}t)$ where
$\omega_{lk}=\omega_l-\omega_k$ and  $A_{lk}$ is a constant
amplitude. Let us compute, at least formally, the averaged system
associated to~\eqref{dynNe:eq} when
$$
0<\Gamma \ll A_{lk}\theta_{lk}\ll \omega_{lk}, \quad \gamma_{lk}\ll
\theta_{lk} .
$$
Consider  the "Pauli matrices" associated to the transition between
$l$ and $k$:
\begin{align*}
& \sigma_x^{lk}= \ket{l}\bra{k} + \ket{k}\bra{l}
 , \quad
\sigma_y^{lk} = -i\ket{l}\bra{k} +i\ket{k}\bra{l}
 \\
&\sigma_z^{lk} = P_l-P_k = \ket{l}\bra{l} - \ket{k}\bra{k}
 ,\quad
 I^{lk} = P_l+P_k =\ket{l}\bra{l} + \ket{k}\bra{k}
\end{align*}
For each $l\neq k$, we have the usual relations:
$$
(\sigma_x^{lk})^2=I^{lk}, \quad \sigma_x^{lk}\sigma_y^{lk} =i
\sigma_z^{lk}, \quad \ldots
$$
For each $j$ and $k\neq j$, $ P_j=\frac{I^{jk}+\sigma_z^{jk}}{2}$.
In the sequel, we use the shortcut notation $\sum_{lk}$ that stands
for $\sum_{1\leq l < k \leq N}$. Thus we have
$$
 u=\sum_{lk} A_{lk}\cos(\omega_{lk}t), \quad
 \mu = \sum_{lk} \theta_{lk} \sigma_x^{lk}
 .
$$
Notice that when $j\neq l$ and $j\neq k$,
$\tr{P_j[\sigma_x^{lk},\hat\rho]}\equiv 0$. When $j=l$, set
$P_l=(I^{lk}+\sigma_z^{lk})/2$ to find
$$
\tr{P_l[\sigma_x^{lk},\hat\rho]} = i\tr{\sigma_y^{lk}\hat\rho}
$$
When $j=k$ we have similarly
$$
\tr{P_k[\sigma_x^{lk},\hat\rho]} = -i\tr{\sigma_y^{lk}\hat\rho} .
$$
Thus~\eqref{dynNe:eq} reads (remember that $P_l-P_k=\sigma_z^{lk}$).
\begin{align*}
   \dotex\hat\rho &=-i[H,\hat\rho]
   -i \sum_{lk}\sum_{l^\prime k^\prime}
   A_{l^\prime k^\prime}\hat\theta_{lk}\cos(\omega_{l^\prime k^\prime}t)
   [\sigma_x^{lk},\hat\rho]
   \\&\qquad
    +\Gamma\sum_{j=1}^N
      \tr{P_j(\rho-\hat\rho)}
      \left(P_j\hat\rho+\hat\rho P_j-2\tr{P_j\hat\rho}\hat\rho\right)
      \\
   \dotex \hat\theta_{lk} &=
  \gamma_{lk}
  \left( \sum_{l^\prime k^\prime}
  A_{l^\prime k^\prime}\cos(\omega_{l^\prime k^\prime}t) \right)
  \tr{\sigma_y^{lk}\hat\rho} (\tr{\sigma_z^{lk} (\rho-\hat\rho)})
\end{align*}

In the inter-action frame,  $\xi=e^{i H t}\rho e^{-i H t}$ and
$\hat\xi=e^{i H t}\hat\rho e^{-i H t}$, we have
\begin{align*}
   \dotex\hat\xi &=
   -i \sum_{lk}\sum_{l^\prime k^\prime}
   A_{l^\prime k^\prime}\hat\theta_{lk}\cos(\omega_{l^\prime k^\prime}t)
   \left[e^{i H t}\sigma_x^{lk}e^{-i H t},\hat\xi\right]
   \\&\qquad
    +\Gamma\sum_{j=1}^N
      \tr{P_j(\xi-\hat\xi)}
      \left(P_j\hat\xi+\hat\xi P_j-2\tr{P_j\hat\xi}\hat\xi\right)
      \\
   \dotex \hat\theta_{lk} &=
  \gamma_{lk}
  \left( \sum_{l^\prime k^\prime}
  A_{l^\prime k^\prime}\cos(\omega_{l^\prime k^\prime}t) \right)
  \tr{e^{i H t}\sigma_y^{lk}e^{-i H t}\hat\xi} \tr{\sigma_z^{lk} (\xi-\hat\xi)}
\end{align*}
since $[P_j,H]=0$. Simple computations show
$$
e^{i H t}\sigma_x^{lk} e^{-i H t} = e^{i \omega_{lk} t
\sigma_z^{lk}}\sigma_x^{lk} = \cos(\omega_{lk}t)\sigma_x^{lk} -
\sin(\omega_{lk}t) \sigma_y^{lk}
$$
and
$$
e^{i H t}\sigma_y^{lk} e^{-i H t} = e^{i \omega_{lk} t
\sigma_z^{lk}}\sigma_y^{lk} = \sin(\omega_{lk}t)\sigma_x^{lk} +
\cos(\omega_{lk}t) \sigma_y^{lk}
$$
For $(l,k)\neq (l^\prime,k^\prime)$,  $|\omega_{lk}|\neq
|\omega_{l^\prime k^\prime}|$. Thus  resonant terms come only from
$(l,k)=(l^\prime,k^\prime)$. The "rotating wave approximation"
of~\eqref{dynNe:eq} reads:
\begin{equation}\label{dynNA1:eq}
 \left\{
 \begin{aligned}
   \dotex\hat\xi =&
   -i \sum_{lk}
   \frac{A_{l k}\hat\theta_{lk}}{2}
   \left[\sigma_x^{lk},\hat\xi\right]
   \\&
    +\Gamma\sum_{j=1}^N
      \tr{P_j(\xi-\hat\xi)}
      \left(P_j\hat\xi+\hat\xi P_j-2\tr{P_j\hat\xi}\hat\xi\right)
      \\
   \dotex \hat\theta_{lk} =&
  \frac{\gamma_{lk} A_{lk}}{2}
  \tr{\sigma_y^{lk}\hat\xi} \tr{\sigma_z^{lk} (\xi-\hat\xi)}
 \end{aligned}
 \right.
\end{equation}
Notice that, instead of using~\eqref{dynNe:eq} as estimator, one can
use  in practice such averaged filter where the large transition
frequencies $\omega_{lk}$ are removed:
\begin{equation}\label{dynNA:eq}
 \left\{
 \begin{aligned}
   \dotex\hat\xi=&
   -i \sum_{lk}
   \frac{A_{lk}\hat\theta_{lk}}{2}
   \left[\sigma_x^{lk},\hat\xi\right]
   \\&
    +\Gamma\sum_{j=1}^N
     \left( y_j(t) -\tr{P_j\hat\xi}\right)
      \left(P_j\hat\xi+\hat\xi P_j-2\tr{P_j\hat\xi}\hat\xi\right)
      \\
   \dotex \hat\theta_{lk} =&
  \frac{\gamma_{lk} A_{lk}}{2}
  \tr{\sigma_y^{lk}\hat\xi} \left( y_l(t)-y_k(t)-\tr{\sigma_z^{lk}\hat\xi}\right)
  .
 \end{aligned}
 \right.
\end{equation}
Let us now generalize  the heuristic argument of
remark~\ref{rem:local} for the stability of~\eqref{dynA1:eq}.
Consider the following function
\begin{equation}\label{eq:lyapN}
 V=\frac{1}{2}
  \sum_{n=1}^{N} \tr{P_n(\hat\xi -\xi)}^2 +
  \sum_{lk}\frac{2(\hat \theta_{lk}-\theta_{lk})^2}{\gamma_{lk}}
  .
\end{equation}
One can easily see that
\begin{multline}\label{eq:gen}
\frac{dV}{dt}=\sum_{lk}\frac{A_{lk}\theta_{lk}}{2}\tr{\sigma_z^{lk}(\xi-\hat\xi)}\tr{\sigma_y^{lk}(\xi-\hat\xi)}\\
-2\Gamma\sum_l\sum_{k<l}\tr{P_k\hat\xi}\tr{P_l\hat\xi}\tr{\sigma_z^{kl}(\xi-\hat\xi)}^2.
\end{multline}
While the second term in~\eqref{eq:gen} is obviously negative, the
first term has no reason to be negative. However, we will show by a
formal argument that (considering some appropriate assumption
concerning the Rabi frequencies) this term can be averaged to zero
and thus can be neglected.

In this aim, consider the real effective Hamitonian:
$$
H_{eff}=\sum_{lk}\frac{A_{lk}\theta_{lk}}{2}\sigma_x^{lk},
$$
and diagonalize it as follows:
$$
H_{eff}=E^\dag \Omega E, \qquad
\Omega=diag(\Omega_1,...,\Omega_N),\qquad E_{lk}\in\RR\quad \forall
l,k.
$$
where $\{\Omega_j\}_{j=1}^N$ are Rabi frequencies of the system.
From now on, we will assume that these Rabi frequencies are
non-degenerate ($\Omega_m\neq \Omega_n$ for $m\neq n$) and moreover
that $\Gamma \ll \Delta_\Omega$ and $\gamma_lk \ll \Delta_\Omega$,
where $\Delta_\Omega=\max_{m\neq n}|\Omega_m-\Omega_n|$.

Now, in analogy with the 2-level case, consider the unitary
transformation
$$
\zeta=U^\dag E \xi E^\dag U, \quad \hat \zeta=U^\dag E \hat\xi
E^\dag U,
$$
where $U(t)=\exp(-it\Omega)$. Under such a transformation $\zeta$ is
trivially constant $\zeta=E\xi_0 E^\dag$. Furthermore, this
transformation also removes the highly oscillating part of $\hat
\xi$, ($|\hat \theta_{lk}-\theta_{lk}|\ll\theta_{mn}$ and $\Gamma\ll
A_{lk}\theta_{lk}$ for all $l,k,m,n$):\small
\begin{multline*}
   \dotex\hat\zeta =
      -i \sum_{lk}
   \frac{A_{lk}(\hat\theta_{lk}-\theta_{lk})}{2}
   \left[U^\dag E \sigma_x^{lk} E^\dag U,\hat\zeta\right]
   \\
    +\Gamma\sum_{j=1}^N
      \tr{U^\dag E P_j E^\dag U(\zeta-\hat\zeta)}
      \left(U^\dag E P_j E^\dag U \hat\zeta+\hat\zeta U^\dag E P_j E^\dag U
      -2\tr{U^\dag E P_j E^\dag U \hat\zeta}\hat\zeta\right).
\end{multline*}\normalsize
Now let us develop the terms in the first part of~\eqref{eq:gen}
using this unitary transformation:\small
\begin{align*}
&\tr{\sigma_z^{lk}(\xi-\hat\xi)}\tr{\sigma_y^{lk}(\xi-\hat\xi)}=\\
&\tr{U^\dag E \sigma_z^{lk} E^\dag U (\zeta-\hat\zeta)} \tr{U^\dag E
\sigma_y^{lk}E^\dag U (\zeta-\hat\zeta)}=\\
&i\left(\sum_{r\neq
s}\exp(i(\Omega_r-\Omega_s)t)(E_{rl}E_{sl}-E_{rk}E_{sk})(\zeta_{sr}-\hat\zeta_{sr})+\sum_{r}(E_{rl}^2-E_{rk}^2)(\zeta_{rr}-\hat\zeta_{rr})\right)\\
&\qquad\qquad\qquad\qquad\left(\sum_{r\neq s}
\exp(i(\Omega_r-\Omega_s)t)(E_{rl}E_{sk}-E_{rk}E_{sl})(\zeta_{sr}-\hat\zeta_{sr})\right).
\end{align*}\normalsize
Developing and removing the highly oscillating terms of frequencies
$\Delta_\Omega$, we find\small
\begin{align*}
\sum_{r\neq s}
&(E_{rl}E_{sl}-E_{rk}E_{sk})(E_{rk}E_{sl}-E_{rl}E_{sk})|\zeta_{sr}-\hat\zeta_{sr}|^2=\\
&\frac{1}{2}\sum_{r\neq s}(
(E_{rl}E_{sl}-E_{rk}E_{sk})(E_{rl}E_{sk}-E_{rk}E_{sl})|\zeta_{rs}-\hat\zeta_{rs}|^2
+\\
&\qquad\qquad(E_{rl}E_{sl}-E_{rk}E_{sk})(E_{sl}E_{rk}-E_{sk}E_{rl})|\zeta_{sr}-\hat\zeta_{sr})^2|=0,
\end{align*}\normalsize
where we have broken the sum into two parts by symmetrizing with
respect to the indices $r$ and $s$.

Even though this argument does not prove the convergence of the
estimator for the multi-level system, it gives a strong reason for
it to be efficient. The simulations of the next section show that
this is effectively the case.

\subsection{Simulations}\label{ssec:simulgen}

Let us check out the performance of this algorithm on two other test
cases: first on a 3-level system and next on the 4-level system also
considered in~\cite{mirrahimi-cocv-05}.

The first test case is given by
$$
H=\begin{pmatrix}
    0&0&0\\
    0&1&0\\
    0&0&3\\
\end{pmatrix} \qquad\text{and}\qquad
\mu=\begin{pmatrix}
    0&1.3&1\\
    1.3&0&-1.5\\
    1&-1.5&0\\
\end{pmatrix}.
$$
We consider a laser field resonant with all the transition
frequencies of the system:
$$
u(t)=A~\left(\sin(t)+\sin(2t)+\sin(3t)\right),
$$
where the laser amplitude $A$ here is chosen to be $0.1$. This
allows us to use the rotating wave approximation (averaging) and
eliminate the Hamiltonian $H_0$ in order to obtain an effective
Hamiltonian $H_{eff}$.\scriptsize
\begin{figure}[h]
\begin{center}
\includegraphics*[width=10 cm, height=5 cm]{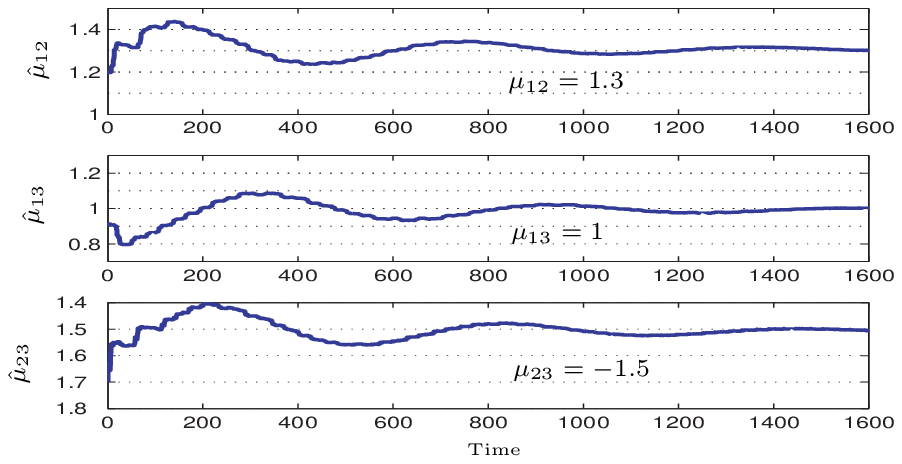}
\caption{This figure illustrates the evolution of the estimator
$\hat\mu$; the estimator gives a very good approximation of the real
dipole moment $\mu$.}\label{fig:3lev-1}
\includegraphics[width=11 cm]{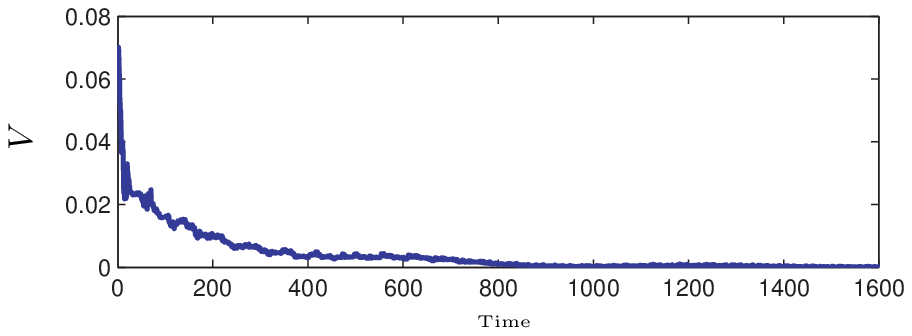}
\caption{The positive definite function $V$ defined
in~\eqref{eq:lyapN} with $\gamma_{lk}=.5$; the function decreases in
average and ends up converging toward 0. } \label{fig:3lev-2}
\end{center}
\end{figure}\normalsize
The Figures~\ref{fig:3lev-1} and~\ref{fig:3lev-2} illustrate the
simulations of the filter equation~\eqref{dynNe:eq} where the
following constants have been used: $\gamma=1$, $\Gamma=.05$,
$A=0.1$. The simulation time $T=1600$ represents around 250 times
the largest period of the system's Hamiltonian $H$. This, however,
represents only about 25 times the largest period of the effective
Hamiltonian $H_{eff}$. Furthermore, the parameter estimator
$\hat\mu$, the state estimator $\hat\rho=\hat\Psi\hat\Psi^\dag$ and
the real state $\rho=\Psi\Psi^\dag$ are initialized at:
$$
\hat\mu(0)=\begin{pmatrix}
    0& 1.2 & .9\\
    1.2& 0 & -1.7\\
    .9& -1.7 &0 \\
\end{pmatrix}\quad,\quad
\hat\Psi(0)=\begin{pmatrix}
    1/\sqrt{14}\\
    2/\sqrt{14}\\
    3/\sqrt{14} \\
\end{pmatrix}\quad,\quad
\Psi(0)=\begin{pmatrix}
    1/\sqrt{30}\\
    2/\sqrt{30}\\
    5/\sqrt{30} \\
\end{pmatrix}.
$$
As one can easily see, the estimator $\hat\mu$ ends up giving a
really good approximation of the true dipole moment in a completely
reasonable time.

Let us now consider the 4-level test case also considered
in~\cite{mirrahimi-cocv-05}. This permits us to have a comparison
between the algorithm provided in this paper and the numerical one
presented in~\cite{mirrahimi-cocv-05}. The physical system is given
by:
$$
H=
    \begin{pmatrix}
    0.0833 & -0.0038 &  -0.0087 &   0.0041 \\
   -0.0038 &  0.0647 &   0.0083 &   0.0038 \\
   -0.0087 &  0.0083 &   0.0036 &  -0.0076 \\
    0.0041 &  0.0038 &  -0.0076 &   0.0357 \\
    \end{pmatrix},
    \quad \mu =
    \begin{pmatrix}
    0 & 5 & -1 &0    \\
    5 & 0 & 6 & -1.5    \\
    -1 & 6 & 0 & 7    \\
    0 & -1.5 & 7 &0    \\
    \end{pmatrix}
$$
Diagonalizing the matrix $H$ yields
$$H=P~D~P^{-1},\qquad
D=
    \begin{pmatrix}
    0 & 0 &  0 &   0 \\
    0 & 0.0365 &   0 &  0 \\
    0 & 0 &   0.0651 &  0 \\
    0 & 0 &   0      &   0.0857 \\
    \end{pmatrix},
    \qquad
P = \exp(\PPP),$$
 where
$$\PPP= \begin{pmatrix}
    0 & 1 &  -1 &   1 \\
    -1 & 0 &   1 &  1 \\
    1 & -1 &   0 &  -1 \\
    -1 & -1 &   1      &   0 \\
    \end{pmatrix}
$$
is an anti-Hermitian matrix. We consider a laser field resonant with
all the transition frequencies of the system:
$$
u(t)=A~\sum_{l=1}^4\sum_{k<l}\sin\left((\lambda_l-\lambda_k)t\right),
$$
where $\lambda_j$ represents the $j$'th eigenvalue of $H$ and the
laser amplitude $A$ is chosen to be $0.01$. \scriptsize
\begin{figure}[h]
\begin{center}
\includegraphics*[width=11 cm, height= 7.6cm]{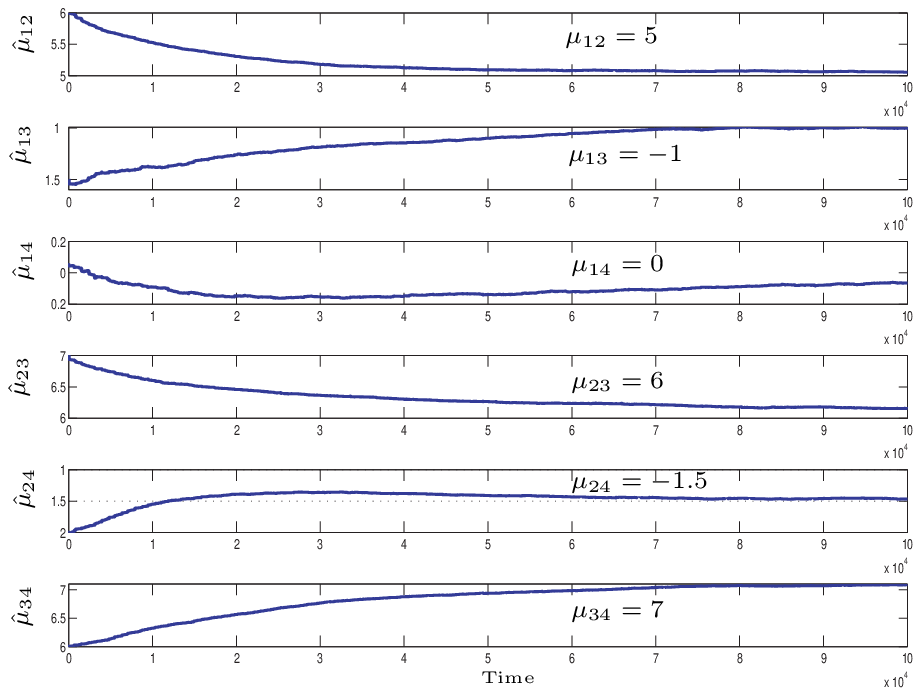}
\caption{This figure illustrates the evolution of the estimator
$\hat\mu$ towards; the estimator gives a very good approximation of
the real dipole moment $\mu$.}\label{fig:4lev-1}
\includegraphics[width=11 cm, height= 3 cm]{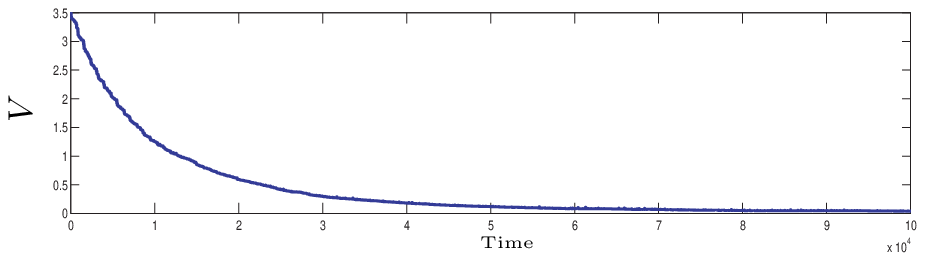}
\caption{The positive definite function $V$ defined
in~\eqref{eq:lyapN} with $\gamma_{lk}=.5$; the function decreases in
average and ends up converging toward 0. } \label{fig:4lev-2}
\end{center}
\end{figure}\normalsize

The Figures~\ref{fig:4lev-1} and~\ref{fig:4lev-2} illustrate the
simulations of the filter equation~\eqref{dynNe:eq} where the
following constants have been used: $\gamma=0.5$, $\Gamma=1$,
$A=0.01$. The simulation time $T=1e+05$ represents around 320 times
the largest period of the system's Hamiltonian $H$. This represents
about 650 times the largest period of the effective Hamiltonian
$H_{eff}$. Furthermore, the parameter estimator $\hat\mu$, the state
estimator $\hat\rho=\hat\Psi\hat\Psi^\dag$ and the real state
$\rho=\Psi\Psi^\dag$ are initialized at:
$$
\hat\mu(0)=\begin{pmatrix}
    0& 6 & -1.5 & .05\\
    6& 0 & 7 & -2\\
    -1.5& 7 &0 & 6\\
    .05 & -2 & 6 &0\\
\end{pmatrix}\quad,\quad
\hat\Psi(0)=\begin{pmatrix}
    1/\sqrt{30}\\
    2/\sqrt{30}\\
    3/\sqrt{30} \\
    4/\sqrt{30} \\
\end{pmatrix}\quad,\quad
\Psi(0)=\begin{pmatrix}
    1/2\\
    1/2\\
    1/2\\
    1/2\\
\end{pmatrix}.
$$
As one can easily see, the estimator $\hat\mu$ ends up giving a
really good approximation of the true dipole moment.

\section{Robustness}\label{sec:robust}
\begin{figure}[h]
\begin{center}
\includegraphics*[width=11 cm, height=5 cm]{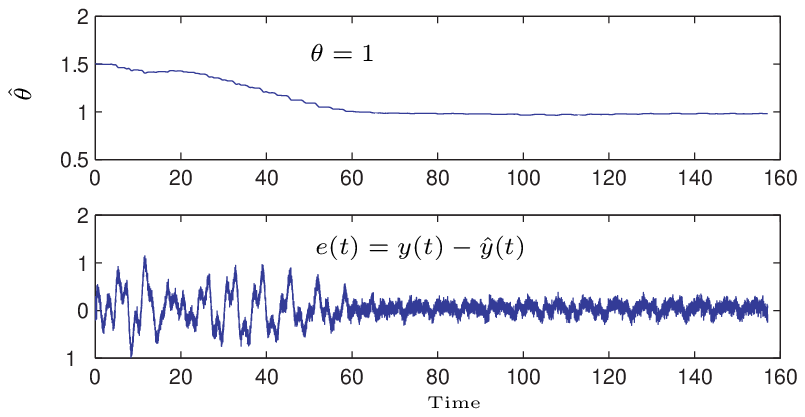}
\caption{ Here we consider the noised measurement of the
form~\eqref{eq:noisemeas}; the first figure illustrates the
robustness, with respect to the uncertainties, in the evolution of
the parameter estimator $\hat\theta$; the second one shows the
robustness in the evolution of the error term
$e(t)$.}\label{fig:rob2lev1}
\includegraphics[width=11 cm, height=3 cm]{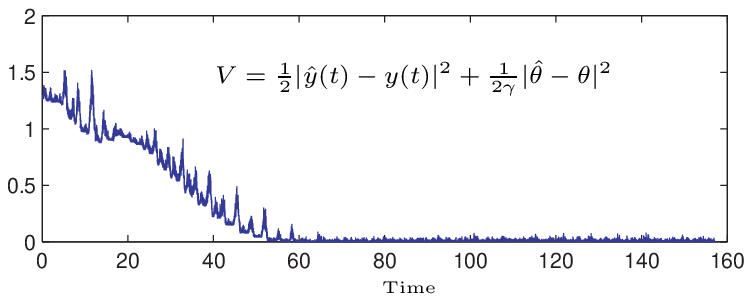}
\caption{The robustness in the evolution of the Lyapunov type
function $V$ with $\gamma=.1$. } \label{fig:rob2lev2}
\end{center}
\end{figure}\normalsize
The laboratory noises are always present and are not negligible.
These noises affect both the output result and the laser field.
Moreover the delay in reading the laboratory output results are
essential and must be taken into account in a faithful model. In
this section, we study the robustness of the algorithm with respect
to all these uncertainties through a number of simulations.
\begin{figure}[t]
\begin{center}
\includegraphics*[width=10 cm, height=5 cm]{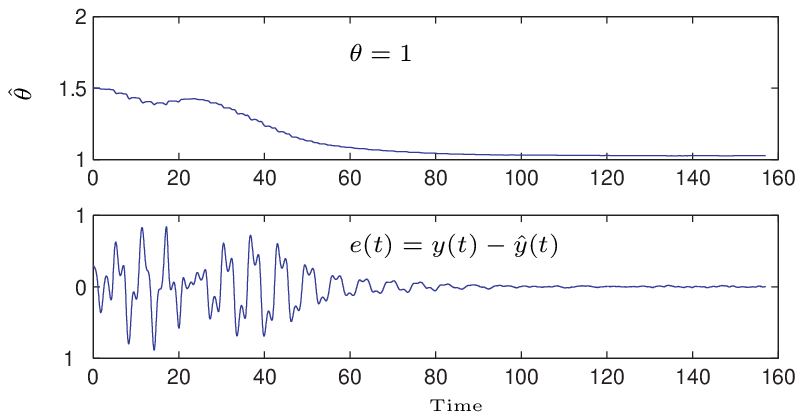}
\caption{ Here we consider the noised control field
of~\eqref{eq:noisecont}; the first figure illustrates the
robustness, with respect to the uncertainties, in the evolution of
the parameter estimator $\hat\theta$; the second one shows the
robustness in the evolution of the error term
$e(t)$.}\label{fig:rob2lev3}
\includegraphics[width=10 cm, height=3cm]{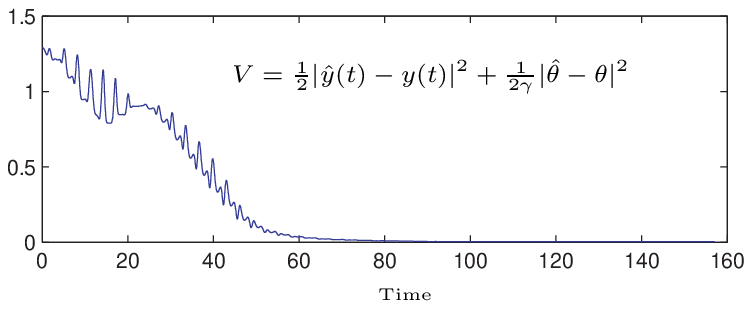}
\caption{The robustness in the evolution of the Lyapunov type
function $V$ with $\gamma=.1$. } \label{fig:rob2lev4}
\end{center}
\end{figure}\normalsize
Concerning the measurement output results, three kind of
uncertainties can be admitted: a delay in reading the output result,
a small additional constant gain and additional non-correlated
noises. The simulations show that the identification algorithm,
presented above, is robust with respect to all these uncertainties.
The simulations of Figures~\ref{fig:rob2lev1} and~\ref{fig:rob2lev2}
show this fact for the 2-level system of Section~\ref{sec:2lev}.
Here we have considered a delay of $0.3$ (about $1/2$ of the natural
period of the system) in reading the measurement results. Moreover,
we have added small additional constant gains and additional
non-correlated gaussian noises:
\begin{equation}\label{eq:noisemeas}
y(t)=\tr{P~\rho(t-0.3)}+0.06+0.07~w,\notag
\end{equation} where
$w$ has a standard normal distribution. Other simulation parameters
are fixed exactly as in the Section~\ref{sec:2lev}.

Regarding the control input, we consider two kind of uncertainties:
a small additive constant gain for the amplitude $A$ of the laser
field and an additional gaussian noise for the laser field. The
simulations of Figures~\ref{fig:rob2lev3} and~\ref{fig:rob2lev4}, on
the 2-level system of Section~\ref{sec:2lev}, show the robustness of
the algorithm with respect to these uncertainties. Here, we have
assumed that the laboratory laser field is noised as follows:
\begin{equation}\label{eq:noisecont}
u(t)=(A+0.03)\sin(t)+0.07~w,
\end{equation}
where $A=1$ as in Section~\ref{sec:2lev} and $w$ is a normal
distribution. Similar simulations concerning the systems of higher
dimensions represent the same kind of robust behavior.
\begin{rem}\label{rem:freq}
One might consider additional uncertainties concerning the
frequencies of the laser field (e.g. small additive constant gains
in the laser frequencies). As it has been discussed in
Remark~\ref{rem:match}, a more realistic estimator in the settings
of our paper is given by the first averaged filter~\eqref{dynA1:eq}
(~\eqref{dynNA:eq} in the case of a multi-level system). The Bohr
frequencies of the system do not appear in this estimator.
Therefore, one can easily check that this averaged estimator
represents a robust behavior with respect to the uncertainties in
the laser frequencies. One only needs these frequencies to be near
enough to the transition frequencies of the system
($\omega_r-\omega\ll A\theta$ in the case of the 2-level system).
\end{rem}

\section{Conclusion}

In this paper, we propose an observer-based method for the
Hamiltonian identification of a quantum system. An intuitive
observer~\eqref{eq:obs} has been considered for the Schr\"{o}dinger
equation~\eqref{eq:2-lev} and has been developed and extended to
give an estimate of the unknown parameters of the system. The
convergence of this method is completely analyzed for a 2-level
case. The multi-level cases have been addressed using heuristic
arguments. Various simulations in different dimensions illustrate
the relevance of the technique for these multi-dimensional systems.
Finally the robustness of the design with respect to different
uncertainties and noises is addressed by simulations on the 2-level
case. Similar robustness results can be noted for multi-dimensional
systems.

In Remark~\ref{rem:freq}, it has been noted that replacing the
estimator~\eqref{dynNe:eq} with the first averaged
version~\eqref{dynNA:eq}, increases considerably the robustness of
the identification result with respect to the frequency
uncertainties.

Such averaged filter represent even more advantages whenever the
settings considered in the paper are valid. In particular, one
increases considerably the robustness with respect to the delay in
the measurement. Indeed, this delay only needs to be much smaller
than the shortest Rabi period of the system. Secondly, the
non-degeneracy assumption for the Rabi  transitions $\Delta_\Omega$
may be removed using a slow modulation of the amplitudes $A_{lk}$.
Finally, one does not really need to have access to the continuous
measurement results $y_j(t)$ (which is lots of information to be
asked in the laboratory settings). In fact, one only needs samples
on the output signal with frequencies much higher than the larger
Rabi frequency. All these advantages seem to highly privilege the
use of the averaged estimator~\eqref{dynNA:eq}.

\appendix
\section{Identifiability}\label{append:identif}

In this appendix, we present the mathematical framework in which the
identification problem can be considered. Moreover, we review
briefly the former work on the identifiability of the considered
system. A well-posedness result which allows us to consider the
identification problem in Section~\ref{sec:gen} will be announced.

The goal is to identify $H=H_0+V$ or/and $\mu$ in system
\begin{equation}\label{eq:main}
i\frac{d}{dt}\Psi=(H_0+V+u(t)\mu)\Psi,\quad \Psi|_{t=0}=\Psi_0,\quad
\|\Psi_0\|_{\HHH}=1,
\end{equation}
when laboratory measurements on some physical observables are
provided. In~\cite{mirrahimi-cocv-05}, two different settings have
been considered in order to characterize the identifiability of such
a system:
\begin{description}
    \item[(S1)] The Hamiltonian $H$ is known and the goal is to
    identify the dipole moment $\mu$. The so-called
    \textit{populations} along the eigenstates $\phi_i$, i.e.
    $p_i=|\her{\phi_i}{\Psi(t)}|^2, i=1,2,...,N$ are measured for
    all instants $t\geq 0$. This is performed with as many control
    amplitudes $u(t)$ as required.
    \item[(S2)] Neither the potential $V$ nor the dipole moment
    $\mu$ are known and the goal is to identify them. Note that,
    by identifying $H$ we mean identifying $V$, as $H_0$ is
    readily known. The eigenvalues of the Hamiltonian $H=H_0+V$
    are also assumed to be known (this assumption is relevant in practice,
    see Remark~\ref{rem:spectroscopy}). Here we measure the
    populations $p_i$ along the states of a canonical basis
    $\{e_i\}_{i=1}^N : p_i=|\her{e_i}{\Psi(t)}|^2, i=1,2,...,N$
    for all instants $t > 0$ and all control amplitudes $u(t)$.
\end{description}
\begin{rem}\label{rem:spectroscopy}
It is relevant in practice to assume that the eigenvalues of the
internal Hamiltonian $H=H_0+V$ are known. In fact the classical
spectroscopy allows for identifying the eigenvalues of the
Hamiltonian and discriminating between two systems that do not share
the same ones. In fact spectroscopy only gives eigenvalue
differences (transition frequencies),  not the absolute values. The
overall unknown additive factor is not seen by the measurements and
has no impact on the identification result.
\end{rem}
In this paper, we have only considered the first setting. An
extension of the technique to the second setting remains to be done
in future work. However, ~\cite{mirrahimi-cocv-05} provides an
identifiability result for this second setting as well.

Here we announce the identifiability result
of~\cite{mirrahimi-cocv-05} concerning the first setting. For a
result in the second setting and also the proof of the result for
the first setting, we refer to~\cite{mirrahimi-cocv-05}.

\begin{thm}\label{thm:identifiability}
Suppose that there exist two dipole moments $\mu_1$ and $\mu_2$,
giving rise to two evolving states $\Psi_1$ and $\Psi_2$
respectively solving
\begin{alignat}{2}
i\dot\Psi_1 &= (H+u(t)\mu_1)\Psi_1,\label{eq:id1}\\
i\dot\Psi_2 &= (H+u(t)\mu_2)\Psi_2,\label{eq:id2}
\end{alignat}
that produce identical observations for all $t\geq 0$ and all fields
$u(t)$:
\begin{equation}\label{eq:meas}
|\her{\Psi_1(t)}{\phi_i}|^2=|\her{\Psi_2(t)}{\phi_i}|^2\qquad
i=1,2,...,N.
\end{equation}
Then under assumptions
\begin{description}
    \item[(A1)] Equation~\eqref{eq:id1} is wavefunction
    controllable~\cite{ramakrishna-et-al-95};
    \item[(A2)] The transitions of the Hamiltonian $H$ are
    non-degenerate: $\lambda_{i_1}-\lambda_{j_1}\neq
    \lambda_{i_2}-\lambda_{j_2}$ for $(i_1,j_1)\neq
    (i_2,j_2)$~\cite{turinici-rabitz-01};
    \item[(A3)] The diagonal part of the dipole moments $\mu_1$
    and $\mu_2$, when written in the eigenbasis of the Hamiltonian
    $H$, is zero: $\bra\phi_i \mu_1\ket \phi_i=\bra\phi_i \mu_2\ket
    \phi_i=0, i=1,2,...,N$;
\end{description}
the two dipole moments are equal within some phase factors
$\{\alpha_i\}_{i=1}^N\subset \RR$ such that:
\begin{equation}\label{eq:incer}
\forall i,j=1,2,...,N, \qquad
(\mu_1)_{ij}=e^{i(\alpha_i-\alpha_j)}(\mu_2)_{ij}.
\end{equation}
\end{thm}
Fore more details and remarks concerning the assumptions and the
result of this theorem we refer to~\cite{mirrahimi-cocv-05}.

\end{document}